\documentclass[twocolumn,secnumarabic,amssymb,nobibnotes,aps,longbibliography,pra,superscriptaddress]{revtex4-1}

\usepackage{graphicx,psfrag,xcolor,hyperref}
\usepackage{amssymb,latexsym,mathrsfs,amsmath}
\usepackage{verbatim}
\usepackage{enumitem}
\usepackage[utf8]{inputenc}

\begin{document}

\def\be{\begin{equation}}
\def\ee{\end{equation}}
\def\bea{\begin{eqnarray}}
\def\eea{\end{eqnarray}}

\title{Information Theoretical Limits for Quantum Optimal Control Solutions: Error Scaling of Noisy Control Channels}


\author{Matthias M. M\"{u}ller}
\email{ma.mueller@fz-juelich.de}
\affiliation{Forschungszentrum J\"{u}lich GmbH, Peter Gr\"{u}nberg Institute - Quantum Control (PGI-8), D-52425 Jülich Germany}

\author{Stefano Gherardini}
\affiliation{CNR-INO, Area Science Park, Basovizza, I-34149 Trieste, Italy}
\affiliation{Department of Physics \& Astronomy \& LENS, University of Florence, I-50019 Sesto Fiorentino, Italy}

\author{Tommaso Calarco}
\affiliation{Forschungszentrum J\"{u}lich GmbH, Peter Gr\"{u}nberg Institute - Quantum Control (PGI-8), D-52425 Jülich Germany}
\affiliation{Institute for Theoretical Physics, University of Cologne, D-50937 Cologne, Germany}

\author{Simone Montangero}
\affiliation{Department of Physics \& Astronomy "G. Galilei", University of Padua, and with INFN Sezione di Padova, I-35131 Padua, Italy}

\author{Filippo Caruso}
\affiliation{Department of Physics \& Astronomy \& LENS, University of Florence, I-50019 Sesto Fiorentino, Italy}

\begin{abstract}
Accurate manipulations of an open quantum system require a deep knowledge of its controllability properties and the information content of the implemented control fields. By using tools of information and quantum optimal control theory, we provide analytical bounds (information-time bounds) to characterize our capability to control the system when subject to arbitrary sources of noise. Moreover, since the presence of an external noise field induces open quantum system dynamics, we also show that the results provided by the information-time bounds are in very good agreement with the Kofman-Kurizki universal formula describing decoherence processes. Finally, we numerically test the scaling of the control accuracy as a function of the noise parameters, by means of the dressed chopped random basis (dCRAB) algorithm for quantum optimal control.
\end{abstract}

\maketitle

\section{Introduction}

A quantum system is defined as open when it interacts with other systems or an environment with several degrees of freedom. Such an interaction radically changes the dynamics of the system, e.g., making the dynamics propagator a non-unitary operator~\cite{PetruccioneBook,CarusoRMP2014}. The semi-classical effects of the interaction between a quantum system and the external environment can be reliably modeled by adding stochastic noise fields on the operators that govern the Hamiltonian evolution of the quantum system. As a result, the evolution of the system turns out to be described by a stochastic Schr\"{o}dinger equation generating as solutions the functionals that are usually known as quantum trajectories~\cite{DaviesCMP1970,Wiseman1996,RivasBook2012,GheraNJP2016,MullerADP2017,RossiPRA2017}. 
An example of a relevant noise source -- especially in biological, solid-state systems and atomic clocks -- is provided by the $1/f$ noise~\cite{WeissmanRMP1988,Book-Pink-noise,Paladino2014,BorregaardPRL2013,NorciaScience2019,DorscherCommPhys2020}. This kind of noise, which decreases as a function of the frequency with a hyperbolic trend, is usually responsible for destroying the phase coherence terms of quantum operations.
In order to amend it, the prediction of the spectral properties of the external noise sources~\cite{MullerPRA2016,MullerSciRep2016,DegenReview2017,SzankowskiJPCM2017}, e.g., by means of quantum estimation methods~\cite{ParisBookEstimation}, is going to play a key role to enhance the performance of quantum technology devices~\cite{RonnowScience2014}.

To tackle efficiently the problem of steering a noisy quantum system in a desired way, a variety of solutions have been introduced. A widely used tool to mitigate the detrimental effect of the interaction of a quantum system with its environment is provided by the dynamical decoupling (DD) of the system from the environment~\cite{ViolaPRL1999}. This approach allows to enhance or suppress certain desired interaction modes~\cite{KofmanNAT2000,KofmanPRL2001,GordonJPB2007,Biercuk2011,GreenNJP2013,Paz-Silva2014} and the protocols can also be optimized by analytic or numeric optimization algorithms~\cite{GordonPRL2008,Clausen2012,MullerSciRep2018,Poggiali2018}. A different approach, also followed in this article, is the extension of quantum optimal control (QOC) methods~\cite{Jurdjevic1996,DAlessandro2007,Brif2010, Glaser2015,Mueller2022} to open quantum systems~\cite{Koch2016,Stefanatos2004,Schmidt2011,Mukherjee2013,Hoyer2014,Kallush2014,Stefanatos2014,Pawela2015,Reich2015,Mukherjee2015,Lovecchio2016}. Commonly used algorithms for QOC include Krotov~\cite{Konnov1999} gradient ascent pulse engineering (GRAPE)~\cite{Khaneja2005} and optimization in the chopped random basis (CRAB)~\cite{Doria2011,Caneva2011,Mueller2022}, e.g., using the dressed CRAB (dCRAB) algorithm~\cite{Rach2015}. In this regard, it is worth mentioning that the CRAB algorithm works by expanding the control field onto a truncated basis and then optimizing the coefficients of the expansion through a gradient-free minimization. The dCRAB algorithm uses additional basis changes to ensure guaranteed convergence to the optimal solution of the considered control problem whenever this is guaranteed by gradient-based methods.

The smoothness and bandwidth of the QOC solution is an important constraint that can be included in the QOC algorithm. The CRAB/dCRAB algorithm naturally produces a bandwidth-limited solution that is given by an ansatz implementing a basis with the desired properties. Such a spectrally-limited control has been achieved also with gradient-based QOC algorithms such as Krotov~\cite{Reich2014} and GRAPE~\cite{Motzoi2011}. Alternatively, also the CRAB-ansatz can be combined with gradient optimization. The gradient-based algorithms of the CRAB family are known as Gradient Optimization of AnalyTic controls (GOAT)~\cite{Machnes2018}, Gradient Ascent in Function Space (GRAFS)~\cite{Lucarelli2018} or Gradient Optimization Using Parametrization (GROUP)~\cite{Soerensen2018} and a comparison among these algorithms is undertaken in Ref.~\cite{Soerensen2018}. In this paper, we employ the dCRAB algorithm~\cite{Doria2011,Caneva2011,Rach2015,Mueller2022}, which has been successfully demonstrated experimentally on different quantum platforms~\cite{Lovecchio2016,Frank2014,Omran2019}, including the use of closed-loop optimization~\cite{Rosi2013,Frank2017,Heck2018,WeidnerPRL2018,Oshnik2022,Marshall2021}.

An important limit for any quantum control solution is given by the time-energy bound known as quantum speed limit (QSL)~\cite{Margolus1998,Deffner2017} that fundamentally reflects the time-energy uncertainty relation. The latter can be formulated also for open quantum systems~\cite{Deffner2013} and its effects can be observed in QOC applications when the pulse operation time approaches the theoretical QSL~\cite{Caneva2009}.

Another bound describing QOC performance is provided by the information content of the control pulse: in Ref.~\cite{Moore2012} the so-called ``$2M-2$-rule'' (with $M$ being the dimension of the quantum system) for the degrees of freedom of the control field was confirmed by numerically studying different QOC problems, and later this was also found for quantum many-body systems~\cite{Caneva2014}.  Specifically, for a perfect state transfer in an $M$-dimensional system, $2M-2$ real coefficients of the state vector have to be brought to the target value, and the control field has to 
contain at least the same number ($2M-2$) of free parameters to fulfill this task. Arguments based on information theory already allowed to quantify the information content of a control field, by introducing bounds on the control error and minimum pulse operation time~\cite{Lloyd1}.
In fact, by using the tools from classical and quantum information theory~\cite{CarusoRMP2014,BookCover,Shannon1948,Shannon1949}, one can consider the control pulse as a communication signal whose correct reception is equivalent to the achievement of the desired control task with suitably small error. The same reasoning applies also when the control over the quantum system is achieved through another quantum system (fully-quantum control) that is denoted as \textit{quantum controller} and interacts with the quantum system to be controlled through a quantum communication channel~\cite{Wu2015,Gherardini2020}. In this regard, it is worth noting that any communication channel is subject to (usually correlated) external noise sources that have the effect of degrading the transmitted signals or losing information packets. However, the modeling of such signal degradation has allowed to understand the processes underlying system-noise field interactions, as for example in \cite{BylickaSciRep2014} where correlated quantum dynamics (also leading to non-Markovian evolutions) has been analyzed with an information theory perspective.

In this paper, we assume that noise sources affect the internal Hamiltonian of the quantum system under analysis, governing the coherent part of its dynamical evolution. This means that the interaction between the quantum system and the external environment leads to extra non-deterministic terms proportional to an external stochastic field, modeled by a stochastic Schr\"{o}dinger equation. Thus, after deriving the master equation that describes the mean stochastic dynamics of the system, averaged over the noise realizations, we evaluate the channel capacity associated to the considered optimal control problem, whereby optimized pulses control a noisy quantum system. According to Shannon's theorems~\cite{Shannon1948,Shannon1949}, the values of a channel capacity in the frequency domain depend on the power spectral density of the noise affecting the channel, as well as the error in reaching a desired quantum state. Thus, in this paper the relations between these quantities are quantitatively characterized, both analytically and by means of numerical simulations, and we show that they are in agreement with the Kofman-Kurizki universal formula for decoherence in quantum processes~\cite{KofmanNAT2000,KofmanPRL2001}.

\section{Control Problem}
\subsection{System dynamics}\label{sec:II.1}

We consider a quantum system described by a state $\rho(t)$, where the time evolution is given by the stochastic Schr\"{o}dinger equation (SSE)
\be
\dot{\rho}(t) = -i[H(t),\rho(t)]\,,\label{eq:Schroedinger-equation}
\ee
with $\hbar$, reduced Planck constant, set to $1$. Here, the Hamiltonian
\be
H(t) = H_d + f(t)H_c+\xi(t)H_n \label{eq:Hamiltonian}
\ee
consists of a drift operator $H_d$, a control term $f(t)H_c$ (with control operator $H_c$ and control field $f(t)$) and the stochastic noise $\xi(t)H_n$, where $H_n$ is a fixed operator and $\xi(t)$ is a stochastic field. The drift and the control operators together can be interpreted as the system Hamiltonian $H_s(t)=H_d+f(t)H_c$, while the stochastic noise term can be interpreted as a perturbation $H_p(t)=\xi(t) H_n$ to which the quantum system is subjected, and its time evolution becomes stochastic. In principle, one could also have multiple control fields and multiple noise fields. Here, we do not treat this case specifically but we will discuss at specific points in the paper how the formulas we derive could be generalized in this regard.

\subsection{Control objective}

The goal of the quantum optimal control problem is to find the optimal control pulse $\hat{f}(t)$ able to drive the quantum system from the initial state $\rho(0)$ to the target state $\rho_T$ at the final time $T$. The optimization problem that provides $\hat{f}(t)$ does not necessarily have an exact solution. This necessarily entails a non-zero control error $\varepsilon$, meaning that the optimal control pulse does not perfectly drive the system to the target state $\rho_T$ but to a final state $\rho(T)$ in the $\varepsilon$-ball around the target state~\cite{Lloyd1}. The control error $\varepsilon$ is commonly expressed as a function of the Uhlmann fidelity $\mathfrak{F}(\rho_T,\rho(T))$ 
between the target and the final states~\cite{Uhlmann1976}. In our context, the latter is slightly modified as 
\be
\mathfrak{F}(\rho_T,\langle\rho(T)\rangle)\equiv\left(\text{Tr}\sqrt{\sqrt{\rho_T}\,\langle\rho(T)\rangle\sqrt{\rho_T}}\right)^{2}
\ee
that compares the target state $\rho_T$ and the average final state $\langle\rho(T)\rangle$ such that $\varepsilon \equiv 1 - \mathfrak{F}(\rho_T,\langle\rho(T)\rangle)$, where the averaging $\langle\cdot\rangle$ of $\rho(T)$ is taken over the statistics of the noise field $\xi(t)$.

\subsection{Optimization algorithm}

In the example section we will study how well for specific control problems the control objective can be achieved by means of QOC. In particular, we will employ the dCRAB algorithm~\cite{Rach2015} to solve the optimization problem. This algorithm makes an ansatz for the optimal solution of the form
\begin{equation}\label{eq:CRAB1}
f(t)=\sum_{i=1}^{N_c}c_{i}f_{i}(t)\,,
\end{equation}
where the basis functions $f_i(t)$ span a subspace of the (infinite-dimensional) space that defines the unconstrained control field, and the optimization is then performed on this subspace of smaller dimension. This can in principle be done by any direct search method: in this work we use the Nelder-Mead simplex algorithm~\cite{Nelder1965} to find the optimal set of coefficients $c_{i}$ $(i=1,\dots\, N_c)$. To exploit the usually advantageous properties of the control landscape~\cite{Brif2010,Chakrabarti2007} that could be distorted by the finite dimensional expansion, after convergence of the direct search method a basis change can be introduced and the optimization be continued in an iterative way:
\begin{eqnarray}\label{eq:dCRAB}
 f^j(t)=c_0^j f^{j-1}(t)+\sum_{i=1}^{N_c}c_i^j f_i^j(t)\,,
\end{eqnarray}
where $f_i^j(t)$ are the new basis functions, and $f^{j-1}(t)$ is the optimal solution from the ($j-1$)th iteration. The coefficient $c_0^j$ allows the optimization to move along the direction of the old pulse, while the coefficients $c_i^j$ ($i=1,\dots,N_c$) allow for the movement along the new search directions $f_i^{j}(t)$.

If we choose the basis functions $f_i^j(t)$ to be trigonometric functions with random frequencies $\omega_i^j$ and these frequencies are chosen to lie in an interval of a specified bandwidth, we can incorporate in a natural way a bandwidth constraint for $f(t)$. As we will see below, we will also need to fix the pulse power to a desired value by simply rescaling the coefficients $c_i^j$ 
before computing the system evolution.

\section{Time-continuous stochastic Schr\"{o}dinger equation and master equation}\label{sec_II}

In this section we derive the master equation obtained by averaging the SSE over the noise realizations. 
In doing this, let us plug the explicit form of the Hamiltonian, Eq.~\eqref{eq:Hamiltonian}, into the Schr\"{o}dinger equation~\eqref{eq:Schroedinger-equation}:
\be\label{eq:diff_equation}
\dot{\rho}(t) = -i[H(t),\rho(t)]
= -i[H_s(t),\rho(t)]-i[H_p(t),\rho(t)],
\ee
where $H_s(t)=H_d+f(t)H_c$ and $H_p(t)=\xi(t) H_n$ as defined in Sec.~\ref{sec:II.1}. The integral form of the initial value problem, which is given by
\be\label{eq:integral_form}
\rho(t)=\rho(0)-i\int_0^t [H(t'),\rho(t')] dt',
\ee
can be re-inserted into the differential equation (\ref{eq:diff_equation}), thus leading to
\bea\label{eq:general-2nd-order}
&\dot{\rho}(t) = -i[H_s(t),\rho(t)]& \nonumber \\
& \displaystyle{ - i\bigg[\xi(t)H_n,\left(\rho(0)-i\int_0^t [H(t'),\rho(t')] dt'\right)\bigg] }.&
\eea

Now, let us assume that the stochastic process $\xi(t)$ follows the distribution $p_t(\xi)$, whereby the mean value and the correlation function of $\xi(t)$ are respectively defined as $\langle\xi(t)\rangle \equiv \int_{\xi}p_t(\xi)\xi d\xi$ and
\be
R_{\xi}(t,t') \equiv \big\langle \xi(t)\xi(t')\big\rangle \equiv \int_{\xi}\int_{\xi'} p_{t,t'}(\xi,\xi')\xi\xi' d\xi d\xi'\,.
\ee
In this way, after averaging over the noise realizations and then considering the noise field $\xi$ and the quantum state $\rho$ as \emph{uncorrelated} stochastic processes, the following master equation is obtained for $\langle \xi(t)\rangle = 0$ (see appendix for more details in its derivation):
\bea\label{eq:Master-equation}
\langle\dot{\rho}(t)\rangle &=& 
-i\left[H_s(t),\langle\rho(t)\rangle\right]  \nonumber\\
&& - \left[H_n,\left[H_n,\int_0^t R_{\xi}(t,t')\langle\rho(t')\rangle dt'\right]\right].
\eea
For the sake of clarity, in the right-hand-side of Eq.\,(\ref{eq:Master-equation}), it is worth observing that the result of taking $\xi$ and $\rho$ as uncorrelated quantities is to allow for the approximation $\int_0^t \langle\xi(t)\xi(t')\rho(t')\rangle dt' \approx \int_0^t \langle\xi(t)\xi(t')\rangle\langle\rho(t')\rangle dt' = \int_0^t R_{\xi}(t,t')\langle\rho(t')\rangle dt'$. As a consequence, the non-coherent part of the master equation (due to noise perturbations) simply depends on the product between the correlation function of the noise, $R_{\xi}$, and the average quantum state $\langle\rho\rangle$. 
Comparable results regarding stochastic dynamics in Hilbert spaces can be also found for instance in Ref.\,\cite{PetruccioneBook,Kallush2014,Stefanatos2014}. 

\section{Noise correlation function and power spectral density}\label{section_noise_features}

The noise correlation function $R_{\xi}(t,t')$ has a central role in characterizing the master equation, Eq.\,(\ref{eq:Master-equation}), obtained by averaging over the noise realizations. In this paper we will investigate control problems involving quantum systems subjected both to white and colored noise sources, under the assumption of $\{\xi(t)\}_{t\in\mathbb{R}}$ being a \textit{weakly stationary} stochastic process. This implies the following properties~\cite{vanKampenBook}:
\begin{enumerate}[label=(\roman*)]
\item 
The mean value of $\xi$ does not depend on the time instant $t$ in which the noise field is sampled, i.e., $\int p_{t}(\xi) \xi d\xi = \int p_{t + t'}(\xi) \xi d\xi$ $\forall t'\in\mathbb{R}$.
\item 
The correlation function is translation-invariant:
$R_{\xi}(t,t')= R_{\xi}(t+t'',t'+t'')$ $\forall t''\in\mathbb{R}$. This implies that we can write $R_{\xi}(t,t') = R_{\xi}(\tau)$ with $\tau \equiv t-t'$.
\item 
The second moment of $\xi$ is finite: $\int p_t(\xi)\xi^2 d\xi < \infty$ $\forall t'\in\mathbb{R}$.
\end{enumerate}
As a result, the master equation (\ref{eq:Master-equation}) reads
\bea\label{eq:stochastic_Schrodinger}
\langle\dot{\rho}(t)\rangle &=& 
-i\left[H_s(t),\langle\rho(t)\rangle\right]\nonumber \\
&&-  \left[H_n,\left[H_n,\int_0^t R_{\xi}(t-t') \langle\rho(t')\rangle dt'\right]\right].
\eea
It is worth noting that the case of non-stationary noise terms, albeit here it is not addressed, can be obtained by considering $\{\xi\}_{t\in\mathbb{R}}$ as a piece-wise stationary stochastic process. From here on, we will omit the symbol $\langle\cdot\rangle$ for the average over noise realizations, unless explicitly stated.

\subsection{Example: Master equation for Gaussian white noise}

Let us assume that $\xi(t)$ is a Gaussian white noise field with zero mean value $\langle\xi(t)\rangle$ and correlation function
\be\label{eq:corr_function_white}
 R_{\xi}(t-t')= 2\gamma\,\delta(t-t')\,,
\ee
which is frequently employed to model thermal noise in electronic devices. Then, the master equation (\ref{eq:stochastic_Schrodinger}) reduces to 
\be
\dot{\rho}(t) = -i[H_s(t),\rho(t)] - \gamma \left[H_n,[H_n,\rho(t)]\right]
\ee
where the parameter $\gamma$ denotes the noise strength. For instance, if we consider a two-level system and $H_n=\frac{1}{2}\sigma_z$ with $\sigma_z \equiv \begin{pmatrix} 1 & 0 \\ 0 & -1 \end{pmatrix}$ Pauli matrix, the master equation becomes
\bea\label{eq:master-equation}
\dot{\rho}(t) = -i[H_s(t),\rho(t)]+ \mathcal{L}(\rho(t))
\eea
with
\bea\label{eq:Lindblad-dephasing}
\mathcal{L}(\rho(t))=\gamma\begin{pmatrix} 0& -\rho_{12}(t)\\ -\rho_{21}(t)&0\end{pmatrix}.
\eea
Hence, also this approach can be employed to get the \textit{microscopic} derivation of the pure-dephasing Gorini-Kossakowski-Sudarshan-Lindblad (GKSL) equation, as conventionally stated in the open quantum systems literature~\cite{PetruccioneBook,CarusoRMP2014,Kallush2014,Gorini1976}.

A microscopic derivation can be carried out also for the decay channel that is modeled by means of the GKSL operator \cite{PetruccioneBook}
\begin{equation}\label{eq:GKSL_op_decay_ch}
\mathcal{L}(\rho(t)) = \gamma\begin{pmatrix}
  \rho_{22}(t) & -\frac{1}{2}\rho_{12}(t) \\
  -\frac{1}{2}\rho_{21}(t) & -\rho_{22}(t)
 \end{pmatrix}.
\end{equation}
To get this result, indeed, let us consider two uncorrelated Gaussian white noise fields $\xi_{x}(t)$ and $\xi_{y}(t)$, acting respectively through the operators 
$\frac{1}{2\sqrt{2}}\sigma_x$ and $\frac{1}{2\sqrt{2}}\sigma_y$ with the Pauli matrices $\sigma_x \equiv \begin{pmatrix} 0 & 1 \\ 1 & 0 \end{pmatrix}$ and $\sigma_y \equiv \begin{pmatrix} 0 & -i \\ i & 0 \end{pmatrix}$. In other terms, $H(t)=H_{s}(t) + \left(\xi_{x}(t)\sigma_x + \xi_{y}(t)\sigma_y\right)/2\sqrt{2}$. The two noise sources are fields with zero mean value and correlation function $R_{\xi_z}(t-t')=2\gamma_{k}\delta(t-t')$ with $k\in\{x,y\}$. Then, we assume that $\gamma_x = \gamma_y = \gamma$. Under these assumptions, the master equation $\dot{\rho}(t) = -i[H_s(t),\rho(t)] - \frac{\gamma}{8}\left(\left[\sigma_x,[\sigma_x,\rho(t)]\right] + \left[\sigma_y,[\sigma_y,\rho(t)]\right]\right)$ reads
\be\label{eq:microscopic_decay_channel}
\dot{\rho}(t) = -i[H_s(t),\rho(t)] + \gamma\begin{pmatrix}
  \rho_{22}(t) - \frac{1}{2} & -\frac{1}{2}\rho_{12}(t) \\
  -\frac{1}{2}\rho_{21}(t) & \frac{1}{2} -\rho_{22}(t)
 \end{pmatrix}
\ee
where the non-coherent part of the right-hand-side of Eq.\,(\ref{eq:microscopic_decay_channel}) is straightforwardly ascribable to the GKSL operator (\ref{eq:GKSL_op_decay_ch}).

\section{Measures of signals information content}

\subsection{Signal-to-noise ratio}\label{sec:signal-to-noise-ratio}

The signal-to-noise ratio is a measure used in statistical signal processing to compare the powers of a desired signal and a noise source~\cite{KaySSPBook}. It is usually characterized in the frequency domain. By properly formalizing the signal-to-noise ratio, one is able to predict the amount of information carried by a specific signal embedded in a noisy environment. In our case, the desired signal is represented by the control Hamiltonian $f(t)H_c$ and in particular by the deterministic time-dependent control field $f(t)$, while the noise is given by the noise Hamiltonian $\xi(t) H_n$, i.e., by the stochastic noise field $\xi(t)$.

As a general definition, the signal-to-noise ratio $S/N$ is defined as the ratio of the power of the signal $S$ and the power of the noise $N$. By denoting with $P_{f}$ the power of the control field $f(t)$ and with $\Sigma_{\xi}$ the power of the stochastic noise field $\xi(t)$, we can define
\begin{equation}
S=P_f ||H_c||^{2}\,\,\,\text{and}\,\,\,N=\Sigma_{\xi} ||H_n||^{2}\,,
\end{equation}
where, without loss of generality, we can set the operator norms to $||H_c||=||H_n||=1$ since the proportionality factors can be absorbed in $f(t)$ and $\xi(t)$. Thus, the signal-to-noise ratio can be written as
\begin{equation}
\label{S_su_N_ratio}
\frac{S}{N} = \frac{P_{f}}{\Sigma_{\xi}}.
\end{equation}
If more than one noise term is present, the noise power $N$ could be generalized to the sum of the power of the single noise terms so that the signal-to-noise ratio is reduced accordingly. Moreover, if more than one control term is present, one could calculate the signal-to-noise ratio for each control term individually.
In Eq.\,(\ref{S_su_N_ratio}), the power $P_{f}$ of the control field is defined as
\begin{equation}
P_{f} \equiv \frac{1}{T}\int_{0}^{T}|f(t)|^{2}dt\,,
\end{equation}
where $T$ is the duration of the control pulse. Using \textit{Parseval's theorem}, the power $P_{f}$ of the control field can be expressed also in the frequency domain, i.e.,
\begin{equation}\label{eq:def_Pf}
P_{f} = \frac{1}{2\pi}\int_{-\infty}^{\infty}\phi_{f}(\omega)\,d\omega\,.
\end{equation}
In Eq.\,(\ref{eq:def_Pf}), $\phi_{f}(\omega)$ denotes the power spectral density of $f(t)$, namely the Fourier transform of its auto-correlation function $R_{f}(\tau) \equiv \int_{0}^{\infty}f(t)f(t - \tau)dt$:
\begin{equation}
\phi_{f}(\omega) = \int_{-\infty}^\infty R_f(\tau)\,e^{-i\omega\tau}d\tau\,.
\end{equation}
Likewise, we can define the power spectral density of the stochastic noise field as
\begin{equation}
\varphi_{\xi}(\omega) = \int_{-\infty}^\infty R_\xi(\tau)\,e^{-i\omega\tau}d\tau
\end{equation}
and consequently the power $\Sigma_\xi$ of the stochastic noise field as
\begin{equation}\label{eq:power_xi}
\Sigma_{\xi} = \frac{1}{2\pi}\int^{\infty}_{-\infty}\varphi_{\xi}(\omega)\,d\omega\,.
\end{equation}
In conclusion, the signal-to-noise ratio can be generally written as
\begin{equation}\label{S_su_N_ratio_2}
\frac{S}{N} = \frac{\displaystyle{\int_{-\infty}^{\infty}\phi_{f}(\omega)\,d\omega}}{\displaystyle{\int_{-\infty}^{\infty}\varphi_{\xi}(\omega)\,d\omega}}\,.
\end{equation}

As an example, let us consider again that $\xi(t)$ is a Gaussian white noise field with zero mean value and correlation function as provided by Eq.\,(\ref{eq:corr_function_white}). Hence, one can find that the power spectral density $\varphi_{\xi}(\omega)$ of the noise field is \emph{constant} and equal to $2\gamma$. Since the integral in Eq.\,(\ref{eq:power_xi}) is performed over all the frequency range $\omega\in(-\infty,+\infty)$, the fact that $\varphi_{\xi}(\omega)$ is constant entails that the power of a white noise is ideally infinite. This evidence is not surprising; a white stochastic process, indeed, is just an idealization for noise processes effectively observed in real systems. Thus, to justify the use of white noise to model a real noise process, let us consider the case of electronic devices affected by thermal noise. Such devices do not work for all the frequency range, but just in the range $\omega\in[-\overline{\omega},+\overline{\omega}]$, where $2\overline{\omega}$ denotes the device bandwidth. In this region, the measured power spectral density of the noise is practically constant, as predicted by the model, and then it decays to zero. Therefore, 
\begin{equation}
\Sigma_{\xi} \approx \frac{\gamma}{\pi}\int^{\overline{\omega}}_{-\overline{\omega}}d\omega = \frac{2}{\pi}\gamma\,\overline{\omega}    
\end{equation}
such that 
\begin{equation}
    \frac{S}{N} \approx \frac{\pi}{2}\frac{||H_c||^{2}}{||H_n||^{2}}\frac{1}{\gamma\,\overline{\omega}\,T}\int_0^T |f(t)|^2 dt \,.
\end{equation}
In section~\ref{sec:VIII} some control problems affected by Gaussian white noise fields will be studied numerically, and the value of $\overline{\omega}$ will depend on the resolution of the numerical time-grid. However, we will circumvent the problem of setting a specific value for $\overline{\omega}$ by considering only the proportionality
\begin{eqnarray}\label{eq:signal-to-noise-ratio-wn}
\frac{S}{N} \propto 
\frac{1}{\gamma}\int_0^T |f(t)|^2 dt
\end{eqnarray} 
that is sufficient to investigate the scaling of the control error as a function of the signal-to-noise-ratio.

\subsection{Channel capacity}
\label{sec:channel_capacity}

According to the Shannon information theory, a channel can asymptotically transmit a message without errors at the maximum rate $\mathcal{C}$, channel capacity \cite{BookCover}. Then, the information (in terms of number of bits) carried by the signal (in our case the control field $f(t)$) is defined as $I_{f}\equiv\mathcal{C}T$, with $T$ the duration of the signal. The channel capacity thus quantifies the information carried by the maximum number $\mathcal{M}$ of distinguishable messages that can be reliably encoded and decoded in a communication procedure through the channel per unit time~\cite{BookCover,CarusoRMP2014}.

In this paper, we specifically investigate how a control field can steer a quantum system, such that its state is transformed in a desired way. In such a QOC problem, the role of the decoded messages is played by the number of distinguishable states that can be reached and the encoded messages are replaced by the set of all admissible control fields $f(t)$. The communication channel then depends on the control landscape related to the specific quantum system dynamics\,\cite{Brif2010,Chakrabarti2007} and on the stochastic noise field. Moreover, the information transferred to the system is provided by the information encoded in the control field $f(t)$ reduced by the detrimental action of the stochastic noise field $\xi(t)$, while the specific control landscape reflects the ability of the quantum system to receive this information. Note that for more than one control field, the total channel capacity could in principle be replaced in first approximation by the sum of the channel capacities for each single contribution to quantify the information transferred to the system. However, the presence of a second control field can also change the features of the control landscape and thus the ability of the quantum system to receive this information.

Let us thus study, from an information theoretical perspective, the expression of the channel capacity ruling the transfer of information from the control apparatus onto the system, by initially neglecting possible limits of the system to employ this information. This will give us an upper bound for the channel capacity of the global channel (i.e., from the control apparatus to the final state of the system dynamics). This intermediate step allows us to apply directly the seminal works of Shannon\,\cite{Shannon1948,Shannon1949} and has the advantage of not being affected by the limitations due to the dimensionality and controllability of the quantum system.

\subsubsection{Noiseless case}

The channel capacity of the noiseless channel is given by \textit{Hartley's law} \cite{Hartley1928}. In such case, what prevents $\mathcal{M}$ (the maximum number of distinguishable messages) from being infinite are (i) a finite dynamic range $\Delta f$ for the amplitude of the encoded signal and (ii) inaccuracies in the signal. The former is related to the magnitude of the control, while the latter to the precision $\delta f$ of the signal (i.e., as generated by the control electronics). Thus, Hartley's law states that in the noiseless case the channel capacity equals to
\begin{equation}
  \mathcal{C} = \Delta\Omega \log_{2}(\mathcal{M}) = \Delta\Omega \log_2 \left(1 + \frac{\Delta f}{\delta f}\right),
\end{equation}
where $\Delta f/\delta f$ is generally denoted as the resolution of the pulse $f(t)$, while
$\Delta\Omega = \omega_{\rm max} - \omega_{\rm min}$ is the channel bandwidth given by the minimal and maximal frequency of the control field, $\omega_{\rm min}$ and $\omega_{\rm max}$, respectively.

\subsubsection{Gaussian noise channels}\label{sub-sec-Gaussian-noise-channels}

Shannon showed~\cite{BookCover,Shannon1948,Shannon1949} that also in the presence of noise, there exists a non-zero value for the channel capacity $\mathcal{C}$ such that the error in transmitting an infinitely long message can be arbitrarily small. In this case, the limitations to the channel capacity are the bandwidth $\Delta\Omega$ and the power of the noise fields. More formally,
the capacity $\mathcal{C}$ of this noisy channel is equal to
\begin{equation}\label{Shannon_law}
 \mathcal{C}= \int_{\omega_{\rm min}}^{\omega_{\rm max}} \log_2 \left(1 + \frac{\phi_{f}(\omega)}{\varphi_{\xi}(\omega)}\right)d\omega\,.
\end{equation}
Therefore, for Gaussian white noise, Eq.~(\ref{Shannon_law}) reduces to
\begin{equation}\label{S-H-theorem}
 \mathcal{C}= \Delta\Omega \log_2 \left(1 + \frac{S}{N}\right)\,,
\end{equation}
where $S/N$ is given by Eq.\,(\ref{S_su_N_ratio_2}). The result of Eq.~(\ref{S-H-theorem}) is commonly known as the Shannon-Hartley theorem.

Note that Eq.\,(\ref{Shannon_law}) holds under the assumption that the noise field is a continuous stochastic process sampled from a Gaussian distribution with known variance. However, as also shown in~\cite{Shannon1949}, the results for Gaussian noise channels are the starting point to characterize classical channel capacities, since we can always provide results -- at least approximately -- for arbitrary noise sources by first considering the Gaussian case.

In what follows we present the main results of our studies. In particular, we provide tight analytical lower-bounds for the error made in controlling a quantum system subject to arbitrary noise fields. For this purpose, we unify the classical Shannon theory for communication in the presence of noise, the information theoretical analysis of QOC as proposed in~\cite{Lloyd1} (but applied also to colored noise fields) and the Kofman-Kurizki decoherence theory for a quantum system coupled to an environment described by a continuum of levels~\cite{KofmanNAT2000,KofmanPRL2001}.

\section{Error limit and time bound}

In this section, we study how the channel capacity of the control problem transforms into a bound for the admissible precision (or error) of the QOC solution, as well as a bound on the required time given the control resources. Such a time reflects an information theoretical speed limit that prohibits steering the quantum system faster at given precision.

Following \cite{Lloyd1}, we first introduce $\mathcal{W}$, which is the set containing all the density operators that are solutions (for all admissible control functions $f(t)$) of the SSE describing the dynamics of the system. $\mathcal{W}$ is also denoted as the set of \textit{reachable states} and depends on the initial state $\rho(0)$. $\mathcal{W}$ has dimension $\dim{\mathcal{W}}\equiv D_{\mathcal{W}}(n)$, which in turn is a function of the dimension $n$ of the system's Hilbert space. Then, we introduce a measure for the \textit{complexity} of the optimal control field, i.e., $D$ that is formally defined as the number of independent degrees of freedom of the control $f(t)$. More practically, $D$ can be equal, for example, to the minimal number of independent bang-bang control pulses\,\cite{ViolaPRL1999} or proportional to the bandwidth or sampling points of $f(t)$. In \cite{Lloyd1} it has been proven that the information content $I_{f}$ carried by the control pulse, with $I_{f}$ proportional to $D$, cannot be smaller than the product of $D_{\mathcal{W}}$ and $-\log_{2}(\varepsilon)$, i.e., $I_{f}\geq -D_{\mathcal{W}}\log_{2}(\varepsilon)$, where for a control error $\varepsilon\in[0,1]$, with $\varepsilon=1-\mathfrak{F}$, $-\log_{2}(\varepsilon)$ represents the self-information of $\varepsilon$. 
It is worth observing that the inequality $I_{f} \geq -D_{\mathcal{W}}\log_{2}(\varepsilon)$ can be rewritten as a lower bound for the control error $\varepsilon$:
\begin{equation}\label{eq:bound_error}
\varepsilon \geq 2^{-\frac{I_{f}}{D_{\mathcal{W}}}}.
\end{equation}
Hence, from $I_{f}\equiv\mathcal{C}T$, one finds that the minimal time needed to reach a given target state $\rho_T$ within $D_{\mathcal{W}}$ with precision $\varepsilon$ is
\begin{equation}\label{eq:bound_T}
T \geq -\frac{D_{\mathcal{W}}}{\mathcal{C}}\log_{2}(\varepsilon).
\end{equation}
Eq.\,(\ref{eq:bound_T}) can be interpreted as follows: The amount of information necessary to solve the QOC problem with precision $\varepsilon$ under a finite channel capacity $\mathcal{C}$ sets a time bound for the evolution of the quantum system.
Note that there can be specific target states that violate this bound (e.g., if the target state is equal to the initial state it can also be reached in zero time) and the bound becomes relevant if the initial and target states are randomly chosen.
Now, for the sake of completeness, we derive the formal expression of the control error bound both in the noiseless case and in the presence of white and colored noise sources.

\subsection{Error bound for a noiseless channel}

In the noiseless case, the channel capacity $\mathcal{C}$ is provided by Hartley's law: $\mathcal{C} = \Delta\Omega\log_{2}(1 + \Delta f/\delta f)$. Hence, $I_{f} = T\Delta\Omega\log_{2}(1 + \Delta f/\delta f) = D\log_{2}(1 + \Delta f/\delta f)$ where $D=\Delta\Omega\,T$, since the bandwidth $\Delta \Omega$ multiplied by the total time reflects the number of degrees of freedom encoded in the control pulse. Accordingly, by substituting the expression of $I_f$ in Eq.\,(\ref{eq:bound_error}), one has
\begin{equation}\label{eq:bound_epsilon_noiseless}
\varepsilon \geq\left(1 + \frac{\Delta f}{\delta f}\right)^{-\frac{D}{D_{\mathcal{W}}}}.
\end{equation}

\subsection{Error bound for Gaussian white noise}

Let us now consider a Gaussian white noise source. According to the Shannon-Hartley theorem, the information carried by the control pulse is $I_{f} = D\log_{2}(1 + S/N)$, where $S/N$ is the signal-to-noise ratio defined above. This entails the following expression for the control error bound:
\be\label{eq:error-limit-wn}
\varepsilon \geq \left(1+\frac{S}{N}\right)^{-\frac{D}{D_{\mathcal{W}}}}.
\ee
Note that both bounds (\ref{eq:bound_epsilon_noiseless}) and (\ref{eq:error-limit-wn}) have been already introduced in Ref.\,\cite{Lloyd1}, in relation to the complexity of a QOC problem and the minimal time needed to accomplish the target control transformation. Now, we want to generalize these bounds to colored noise.

\subsection{Error bound for Gaussian colored noise}

As explained in Section\,\ref{sub-sec-Gaussian-noise-channels}, the capacity $\mathcal{C}$ of a quantum channel perturbed by Gaussian colored noise is given by Eq.\,(\ref{Shannon_law}). Before moving on, it is worth mentioning that for Gaussian colored noise we denote correlated (thus, non-white) random sequences sampled over time by a Gaussian distribution. This means that the information carried by the control pulse is
\be
I_f = T\int_{\omega_{\rm min}}^{\omega_{\rm max}}\log_{2}\left(1 + \frac{\phi_{f}(\omega)}{\varphi_{\xi}(\omega)}\right)d\omega\,.
\ee
Hence, in this case a lower bound for the control error $\varepsilon$ is provided by the relation
\be\label{eq:error-limit-cn}
\log_2 \varepsilon \geq -\frac{T}{D_\mathcal{W}}\int_{\omega_{\rm min}}^{\omega_{\rm max}}\log_{2}
\left(1 + \frac{\phi_{f}(\omega)}{\varphi_{\xi}(\omega)}\right)d\omega\,.
\ee
Let us observe that all the presented lower bounds for $\varepsilon$, as well as the ones that will be introduced later, generally depend on the precision with which optimal pulses are numerically derived and then experimentally implemented on real systems. In addition, to get such a bounds, we have not used so far any specific form to express the control error $\varepsilon$, e.g., whether it is given by $1-\mathfrak{F}$ or $1-\mathfrak{F}^2$ in terms of the fidelity $\mathfrak{F}$ or another metric. For these reasons, to check their validity, we can introduce fitting parameters that relate the obtained control error value with the right-hand-side of the considered lower bound, as done for instance in Ref.\,\cite{Gherardini2020}. Still, we will show in Sections~\ref{sec:VIII} and \ref{sec:IX} that the scaling of the control error $\varepsilon$ as a function of key quantities entering the considered QOC problem corresponds to the bounds derived in this section.

\section{Error bounds from perturbation theory}\label{sec:VII}

Let us recall our interpretation of the Hamiltonian Eq.\,\eqref{eq:Hamiltonian} composed of the unperturbed system (described by $H_s(t)=H_d+f(t)H_c$) and a perturbation (provided by $H_p(t)=\xi(t)H_n$), i.e.,
\begin{equation}\label{eq:perturbation}
H(t)=H_s(t)+H_p(t)\,,
\end{equation}
with the aim to understand how the evolution of the system is influenced by the effect of the perturbation (noise).

If we take only the system Hamiltonian $H_s(t)$ together with the initial state $\rho_s(0)=|\psi_0\rangle\!\langle\psi_0|$, we obtain a Schr\"{o}dinger equation with solution $\rho_s(t)=|\psi(t)\rangle\!\langle\psi(t)|$ for the time evolution of the unperturbed system. Instead, the time evolution of the perturbed system is described by the state $\rho(t)=|\psi(t)\rangle\!\langle\psi(t)|$ that is the solution of the Schr\"{o}dinger equation defined by the perturbed (or full) Hamiltonian $H(t)$ and the initial state $\rho(0)=\rho_s(0)$.

Revisiting Eq.~(\ref{eq:perturbation}) and following Ref.~\cite{Clausen2012}, we can introduce the interaction picture following the unitary evolution $U_s(t)$ of the unperturbed system. In the interaction picture, the Hamiltonian $\tilde{H}_p(t)$ that models the perturbation can be written as
\begin{eqnarray}
	\tilde{H}_p(t) = U_s^{\dagger}(t)H_p(t)U_s(t) \,,
\end{eqnarray} 
while the state evolution of the perturbed system becomes
\begin{eqnarray}
	\tilde{\rho}(t)=U_s^\dagger(t)\rho(t) U_s(t)=|\tilde{\psi}(t)\rangle\!\langle\tilde{\psi}(t)|\,.
\end{eqnarray}
Thus, in the interaction picture, the Schr\"{o}dinger equation is provided by
\begin{eqnarray}
	\dot{\tilde{\rho}}(t) =-i\left[\tilde{H}_p(t),\tilde{\rho}(t)\right],
\end{eqnarray}
with $\tilde{\rho}(0)=\rho(0)$ as $U_s(0)=I$. If the unperturbed system perfectly performs the desired control task, then the control error $\varepsilon_p$ of the perturbed (noisy) system can be expressed as
\begin{eqnarray}\label{eq:perturbation-error}
	\varepsilon_p = 1 - F_p^2 = 1 - |\langle \psi_0|\tilde{\psi}(T)\rangle|^2,
\end{eqnarray}
where the overlap of the perturbed dynamics with respect to the unperturbed dynamics is provided by $F_p^2\equiv \mathfrak{F}(\rho_s(T),\rho(T))= |\langle\psi_s(T)|\psi(T)\rangle|^2=|\langle \psi_0|\tilde{\psi}(T)\rangle|^2$.
We now analyze Eq.~\eqref{eq:perturbation-error} for two different regimes, first in the small noise approximation, and then for stationary colored noise.

\subsection{The small noise approximation}\label{sec:smallnoise}

To estimate the influence of a small noise contribution to the control error we make use of the Dyson expansion of the perturbed state, which up to first order in time yields
\begin{equation}
	|\tilde{\psi}(t)\rangle \approx \Big(1- i \int_0^t\tilde{H}_p(t_1)dt_1\Big) |\psi_0\rangle\,.
\end{equation}
Exploiting this first-order Dyson expansion we thus obtain
\begin{eqnarray}
&\Vert |\psi_s(T)\rangle - |\psi(T)\rangle\Vert = \Vert|\psi_0\rangle - |\tilde \psi(T)\rangle\Vert&\nonumber\\
&\approx \left\Vert |\psi_0\rangle - \left(1 - i \int_0^T\tilde{H}_p(t_1)dt_1 \right)|\psi_0\rangle\right\Vert \leq \Vert H_p\Vert T \,,&
\end{eqnarray}
where the norm $\Vert H_p\Vert$ of the Hamiltonian $H_p$ denotes the maximum values of the corresponding standard operator norm over the time interval $[0,T]$, such that it can be approximated as $\Vert H_p\Vert^2\approx N$. In addition, the operation time $T$ is roughly proportional to the inverse of the norm of the control Hamiltonian $\Vert H_s\Vert^2\approx S$, and thus $T\propto 1/S$, with the result that $\Vert |\psi_s(T)\rangle - |\psi_p(T)\rangle\Vert \lesssim N/S$. Furthermore, with some algebra one has that
\begin{eqnarray}
\Vert |\psi_s(T)\rangle - |\psi(T)\rangle\Vert^2 &=& 2 - 2\,\mathrm{Re}\left(\langle\psi_s(T)|\psi(T)\rangle\right)\\
&\geq& 2- 2F_p\,,
\end{eqnarray}
where $\mathrm{Re}(\cdot)$ stands for the real part of $(\cdot)$.
Thus, for the fidelity $F_p$, it holds that
\begin{eqnarray}\label{eq:Fp_square}
F_p^2 &\geq& \bigg(1-\frac{1}{2} \Vert |\psi_s(T)\rangle - |\psi(T)\rangle\Vert^2 \bigg)^2 \nonumber \\
&\approx& 1 - \Vert |\psi_s(T)\rangle - |\psi(T)\rangle\Vert^2 .
\end{eqnarray}
This entails that, by resolving for the operation error $\varepsilon_p \equiv 1 - F_p^2$, the following upper bound can be obtained: 
\begin{eqnarray}\label{eq:varepsilon_p}
 \varepsilon_p &\lesssim& \Vert H_p\Vert^2 T^2\propto N/S.
\end{eqnarray}
The same result can be achieved also by exploiting the Gronwall's lemma\,\cite{Gronwall}. Specifically, one gets
\begin{equation}
\Vert |\psi_s(T)\rangle - |\psi(T)\rangle\Vert \leq \frac{\Vert H_p\Vert}{\Vert H_s\Vert}\left(
\exp\left\{\int_0^T \Vert H_s(t)\Vert dt \right\} - 1\right)
\end{equation}
where also the norm $\Vert H_s\Vert$ of the control Hamiltonian $H_s$ denotes the maximum value of the standard operator norm over the time interval $[0,T]$. This justifies also the approximation $\Vert H_s\Vert^2\approx S$ introduced above. Hence, by taking $\varepsilon_p \equiv 1 - F_p^2$ (operation error) with $F_p^2 \geq 1 - \Vert |\psi_s(T)\rangle - |\psi(T)\rangle\Vert^2$ [Eq.\,\eqref{eq:Fp_square}], in this case we end-up to the upper bound
\begin{eqnarray}
 \varepsilon_p &\lesssim& \frac{\Vert H_p\Vert^2}{\Vert H_s\Vert^2}\left(\exp \left\{ \int_0^T \Vert H_s(t)\Vert dt \right\}-1\right)^{2}.
\end{eqnarray}
As a result, we can thus state that the operation error $\varepsilon_p$ induced by a noise perturbation (i.e., the noise term in Eq.\,\eqref{eq:Hamiltonian}) scales with $\frac{\Vert H_p\Vert^2}{\Vert H_s\Vert^2} \approx \frac{N}{S}$. 

Finally, if we further assume that without noise ($H_p(t)=0$) one can perfectly control the system such that the final state of the unperturbed system corresponds to the target state $|\psi_s(T)\rangle\!\langle\psi_s(T)| = \rho_{T}$, then the control error is bounded by $\varepsilon\leq\varepsilon_p$ and 
the right-hand-side of Eq.\,\eqref{eq:varepsilon_p} is also an upper bound of the control error $\varepsilon$ for the perturbed (noisy) system. This is consistent with Eq.~\eqref{eq:error-limit-wn} for $S/N \gg 1$.

\subsection{Relation with the Kofman-Kurizki decoherence universal formula}
\label{sec:KK-formula}

The decay of unstable states into a continuum of quantum levels, mimicking a macroscopic reservoir, is well described by the so-called Kofman-Kurizki universal formula~\cite{KofmanPRL2001}. The open dynamics originating from the interaction between a finite-dimensional quantum system and a reservoir leads to decoherence, i.e., the asymptotic loss of the quantum system coherence. We have already seen in Section~\ref{sec_II} and \ref{section_noise_features} that decoherence can be described by modeling the interaction with the reservoir as weak stochastic perturbations. In this regard, the Kofman-Kurizki universal formula, under the hypothesis of weak coupling, predicts how the control of an unstable quantum system can modify its decay rate into the reservoir~\cite{GordonJPB2007,GordonPRL2008,Zwick2016}. The effective value of the decay rate can also be engineered with optimization techniques to suppress or enhance different coupling modes~\cite{GordonPRL2008, Clausen2012, MullerSciRep2018, Poggiali2018}.
Therefore, we phenomenologically expect that the information-time bound of Eq.\,(\ref{eq:bound_T}) for the error scaling and the Kofman-Kurizki universal formula for the quantum system decay rate are related.

To derive the control error based on the Kofma-Kurizki universal formula in this context, we make use of the Dyson expansion of the perturbed state in the interaction picture, which up to the second order in time yields:
\begin{eqnarray}
	|\tilde{\psi}(t)\rangle &\approx \Big(  &1- i \int_0^t\tilde{H}_p(t_1)dt_1 \nonumber\\
	&&- \int_0^t\int_0^{t_1} \tilde{H}_p(t_1)\tilde{H}_p(t_2)dt_1 dt_2 \Big)|\psi_0\rangle \,.\qquad
\end{eqnarray}
Therefore, by performing the Taylor expansion of $|\langle\psi_0|\tilde{\psi}(T)\rangle|^2$ (similarly to what done in Sec.~\ref{sec:smallnoise}), averaging over the noise realizations and substituting $\langle \xi(t)\rangle=0$, one gets
\begin{eqnarray}\label{eq:KK-timedomain}
\varepsilon_p &\approx&
2\,\mathrm{Re}\Big(\int_0^T\int_0^{t_1}\langle\psi_0|\tilde{H}_n(t_1)\tilde{H}_n(t_2)|\psi_0\rangle\nonumber\\
&& \qquad \times\, \langle \xi (t_1)\xi(t_2)\rangle dt_1 dt_2 \Big).
\end{eqnarray}
Thus, if we also introduce the filter function
\begin{eqnarray}
F(\omega) \equiv \frac{1}{\pi}\,\mathrm{Re}\Big( \int_0^T\int_0^{t_1}\langle\psi_0|\tilde{H}_n(t_1)\tilde{H}_n(t_2)
|\psi_0\rangle\nonumber\\
\times\, e^{-i\omega(t_1-t_2)} dt_1 dt_2\Big),
\end{eqnarray}
the right-hand-side of Eq.~\eqref{eq:KK-timedomain} can be written as the overlap of $F(\omega)$ and the noise spectrum $\varphi_{\xi}(\omega)$, i.e.,
\begin{eqnarray}
	\varepsilon_p \approx \bigg|\int_{-\infty}^\infty F(\omega)\varphi_{\xi}(\omega)d\omega\bigg|\,.
\end{eqnarray}

In quantum noise spectroscopy and dynamical decoupling~\cite{ViolaPRL1999,Biercuk2011,GreenNJP2013,Paz-Silva2014,MullerSciRep2018,Yuge2011,Alvarez2011,Bylander2011,Norris2016,Frey2017,Do2019,MartinaArXiv2021,WisePRXQuantum2021}, the control pulse is modulated (by means of a set of pulse modulation functions) to encode information on the noise field in the final state of the system or to minimize the control error. Here, we are interested in the latter case.

To gain more insight into the scaling of the error with the control resources, let us make the following approximations: the filter function $F(\omega)$ is designed to have support only around a central frequency $\omega_c$ (e.g., for bang-bang control, this assumption is feasible since the filter functions have a $\text{sinc}^{2}$-shape, but also for optimally modulated pulses the bandwidth remains small~\cite{GordonPRL2008}) and with a small bandwidth if compared to the power spectral density of the stochastic noise field. Thus, we can make the approximation
\be\label{eq:bound_KK_formula}
 \varepsilon_p \approx C_{0}\,\varphi_{\xi}(\omega_c) \,,
\ee
with $C_{0}$ a constant term depending on the specific choice of the filter function.

We are now in the position to compare this result with the information theoretical limit provided by Eq.\,(\ref{eq:error-limit-cn}). In practical applications, the control pulse is chosen such that the signal-to-noise ratio (evaluated in dB) exceeds a minimum threshold (at least 10-20 dB) so that the noise does not dominate the dynamics and a satisfactory value of the control fidelity $\mathfrak{F}$ (e.g., more than $90\%$) can be achieved. We thus assume that $\phi_{f}(\omega)/\varphi_{\xi}(\omega)\gg 1$ $\forall\omega$ 
such that $\log_2 \left(1 + \phi_{f}(\omega)/\varphi_{\xi}(\omega)\right) \approx \log_2 \left(\phi_{f}(\omega)/\varphi_{\xi}(\omega)\right)$. As a result,
\begin{equation}
\log_{2}\varepsilon \gtrsim -\frac{T}{D_{\mathcal{W}}}\int_{\omega_{\rm min}}^{\omega_{\rm max}} \log_{2}\left(\frac{\phi_{f}(\omega)}{\varphi_{\xi}(\omega)}\right)d\omega\,.
\end{equation}
Then, let us again consider that the bandwidth of the control field is smaller than the frequency variation of the power spectral density $\varphi_{\xi}(\omega)$ of the noise field (this assumption is the same that we have made for the filter function $F(\omega)$ to derive the bound in Eq.\,(\ref{eq:bound_KK_formula}). Since $f(t)$ enters the dynamics only through the filter function $F(\omega)$ that contains the control degrees of freedom (see also the arguments in section~\ref{sec:channel_capacity}), the integral
$\int_{\omega}\log_{2}\left(\phi_{f}(\omega)/\varphi_{\xi}(\omega)\right)d\omega$ can be approximated by its integrand $\Delta\Omega \log_{2}\left(C_1/\varphi_{\xi}(\omega_c)\right)$, evaluated at the central frequency $\omega_c$ of the effective pulse modulation.
The constant $C_1$ depends on the precise value and shape of $\phi_f(\omega)$. As a consequence,
\be\label{eq:bound-it-cn}
\varepsilon \gtrsim \left(\frac{\varphi_{\xi}(\omega_c)}{C_1}\right)^{\frac{\Delta\Omega\, T}{D_{\mathcal{W}}}} \approx C_{2}\,\varphi_{\xi}(\omega_c)\,,
\ee
with $D=\Delta\Omega\,T$ chosen to be approximately equal to $D_{\mathcal{W}}$ and such that $C_{2} \approx 1/C_1$ is a constant. The assumption $D\approx D_{\mathcal{W}}$ is justified by considering that for a value of $D$ smaller than $D_{\mathcal{W}}$, even without noise, one may not have full control over the quantum system; while, for a value of $D$ substantially larger than $D_{\mathcal{W}}$, one potentially has to spread out the power of the control pulse over a larger frequency range (with the consequence of a power spectral density with a lower amplitude in that range), such that the signal-to-noise ratio becomes less favorable.

We have thus shown that, under reasonable assumptions, the information theoretical bound of Eq.\,(\ref{eq:error-limit-cn}) leads to the same scaling of the error as a function of the noise as the one obtained in Eq.\,(\ref{eq:bound_KK_formula}), by starting from the Kofman-Kurizki universal decoherence formula.

\subsubsection*{Example: 1/f-noise}

As an example, let us here consider that the noise field is sampled by a $1/f$ distribution, commonly used to model statistical fluctuations of a wide number of physical and biological systems \cite{WeissmanRMP1988,Book-Pink-noise,Paladino2014}. As explained in Refs.\,\cite{WeissmanRMP1988,Book-Pink-noise, Paladino2014}, the power spectral density of the $1/f^\alpha$-noise decays as a power-law with the frequency, i.e., $\varphi_{\xi}(\omega)\propto 1/\omega^\alpha$ where the exponent $\alpha$ typically belongs to the range $0<\alpha<2$ (see also Refs.\,\cite{Wiseman1996}). The case of $\alpha=1$ is also called pink noise ($1/f$-noise). Thus, the natural choice in this case is to control the system at high frequency. According to Eq.~\eqref{eq:bound_KK_formula}, when moving to higher frequencies, the error $\varepsilon$ decays as $\varepsilon \approx C_{0}/\omega_c^\alpha$. Likewise, if we consider the error-bound arising from the Shannon-Hartley theorem, Eq.~\eqref{eq:bound-it-cn}, we find that $\varepsilon \gtrsim C_{2}/\omega_c^\alpha$, thus confirming the result.

\section{Examples with Gaussian white noise}\label{sec:VIII}

In this section we numerically examine the state-transfer control of a two-level system subject to Gaussian white noise. The resulting dynamics is provided by the master equation of Eq.~\eqref{eq:master-equation}.
In subsection \ref{sec:dephasing-channel} the control error $\varepsilon$ is studied as a function of the signal-to-noise ratio for a system subject to dephasing, while in section \ref{sec:decay-channel} we study $\varepsilon$ as a function of the control bandwidth for a system under the influence of a decay term. In both cases, our aim is to extract the scaling of the control error from the information theoretical bound for white noise fields as given by Eq.~\eqref{eq:error-limit-wn}, and then to confirm the scaling by fitting the bound to the numerical data.

\subsection{Dephasing channel}\label{sec:dephasing-channel}

\begin{figure}[t]
	\centering
	\includegraphics[width=0.49\textwidth]{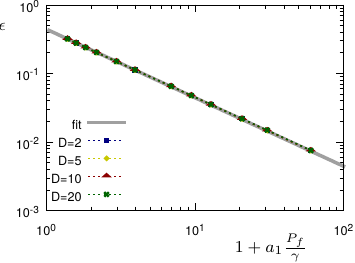}
	\caption{Dephasing channel (Sec.~\ref{sec:dephasing-channel}): Control error scaling with respect to the signal-to-noise ratio $S/N\propto P_f/\gamma$ for different values of $D$. For the parameters involved in the simulations, we have chosen the values $T=1$, $\omega_y=2\pi\times 0.5$, $\omega_z=2\pi\times 0.5$. Then, in each iteration of the optimization, the noise strength $\gamma$ is chosen as a function of the control pulse power $P_{f}$ to set the value of $S/N$. The data points plotted in the figure represent the minimum error from 10 optimization runs for each value of $D$, namely ($D=2$, blue squares),  ($D=5$, yellow circles),  ($D=10$, red triangles), and  ($D=20$, green crosses). Finally, the fit curve (grey line, `fit') is the result of the fit provided by Eq.~\eqref{eq:model-dephasing} in the main text, with $a_1=0.0754$ and $b_1=0.443$.}
	\label{fig:SN-scan}
\end{figure}

Let us consider a two-level system with Hamiltonian $H_s(t)$ given by
\begin{equation}
H_1(t) \equiv f(t)\sigma_x + \omega_y \sigma_y + \omega_z\sigma_z \,,
\end{equation}
where $f(t)$ is the control field, $T = 1$ and $\omega_y = \omega_z = 2\pi\times 0.5$. Then, we model the dephasing acting on the quantum system via the GKSL operator of Eq.~\eqref{eq:Lindblad-dephasing}. To fix the signal-to-noise ratio to the desired value, at each optimization iteration we calculate the power of the control field $P_f=\frac{1}{T}\int_0^T |f(t)|^2 dt$, and we set the value of the noise strength $\gamma$ such that the signal-to-noise ratio
$S/N \propto \frac{P_f}{\gamma}$ (see Eq.\,\eqref{eq:signal-to-noise-ratio-wn}) is constant during the optimization. This corresponds to a physical model where the noise is induced by a frequency instability of the control pulse, and the amplitude of the instability is proportional to the pulse strength.

Our aim is to control the transfer from the initial state $\rho(0)=|0\rangle\!\langle 0|$ to the target state $\rho_T=|1\rangle\!\langle 1|$, by imposing a fixed bandwidth $\Delta\Omega$ of the control field $f(t)$, where $|0\rangle$ and $|1\rangle$ are the eigenstates of $\sigma_z$. In the simulations $D_{\mathcal{W}} = 2$, where in this case $D_{\mathcal{W}}$ represents the dimension of the reachable qubit (pure) states with disregard of the global phase, and $D = \frac{\Delta\Omega}{2\pi}T$. Fig.~\ref{fig:SN-scan} shows the resulting control error obtained by dCRAB optimization with control bandwidth $\Delta\Omega \in 2\pi\times \{2,5,10,20\}$.
We can observe that, in this example, the optimization error does not depend on the control bandwidth, and instead the control error $\varepsilon$ is dominated by the signal-to-noise ratio $S/N$. From Eq.~\eqref{eq:error-limit-wn} we thus expect that
\begin{eqnarray}\label{eq:model-dephasing}
\varepsilon_1 \gtrsim
b_1 \frac{1}{1+a_1\frac{P_f}{\gamma}}
\end{eqnarray}
up to some constants $a_1$ and $b_1$, where $a_1$ determines $S/N=a_1 P_f/\gamma$ according to the reasoning in section~\ref{sec:signal-to-noise-ratio}, and $D/D_\mathcal{W}\approx 1$ for $D\geq 2$ 
to saturate the information transferred from the control pulse onto the quantum system. To verify the model of Eq.~\eqref{eq:model-dephasing} with the numerical data, we take the data for $D=2$ and then we fit the model 
to the control errors obtained from the numerical simulations. The same fit is done also for the alternative model $\varepsilon_2= b_2\exp(-a_2\frac{P_f}{\gamma})+c_2$. If the fit for the first model produces a lower root-mean-square of the residuals (RMS), then this effectively proves our prediction for the scaling of the error (i.e., a power-law instead of an exponential scaling).
In Fig.~\ref{fig:SN-scan} we show the fit of $\varepsilon_1$, for which we obtain a RMS of $0.00094$, with $a_1=0.0745$ and $b_1=0.443$. Instead, for $\varepsilon_2$ we have a RMS of $0.014$, which thus confirms our model. It is worth observing that this finding is consistent with the perturbative treatment of the quantum system's dynamics described in Sec.~\ref{sec:smallnoise}. We can thus conclude that, in general, increasing $D$ above the value of $D_{\mathcal{W}}$ saturates the information that can be transferred onto the system and the original exponent of the bound \eqref{eq:error-limit-wn} can be approximated by $-D/D_{\mathcal{W}}\approx -1$. We have thus seen that in this numerical example the control error follows the information theoretical bound for Gaussian white noise, Eq.~\eqref{eq:error-limit-wn}, as well as the prediction from the perturbative small-noise approximation of Sec.~\ref{sec:smallnoise}\,.

\subsection{Decay channel}\label{sec:decay-channel}

As a second example, we consider a two-level quantum system, where the Hamiltonian $H_s(t)$ is provided by
\begin{eqnarray}
H_2(t) \equiv f(t)\sigma_x + \omega_z\sigma_z\,,
\end{eqnarray}
and a decay term described by the GKSL operator of Eq.~\eqref{eq:GKSL_op_decay_ch}. For the numerical simulations we set $T=1$, $\omega_z=2\pi\times 1$, and $\gamma=2\pi\times 0.4$. Then, we fix the signal power $S$, such that $\int_0^T |f(t)|^2 dt = (2\pi\times 4)^2$, and thus the signal-to-noise ratio by following Eq.~\eqref{eq:signal-to-noise-ratio-wn}.

\begin{figure}[t]
   \centering
   \includegraphics[width=0.49\textwidth]{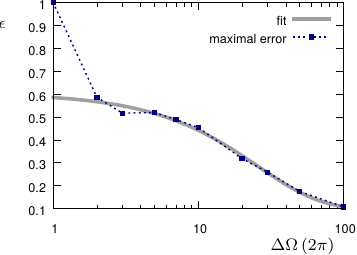}
   \caption{Decay channel (Sec.~\ref{sec:decay-channel}): Control error scaling with respect to the maximum bandwidth $\Delta\Omega$. The control error $\varepsilon$ is reduced when a higher bandwidth allows faster transitions between the two states. The figure shows the maximal control error over 100 random instances of input and target states (blue squares). If the bandwidth is too low, then we do not have any control over the quantum system and the maximal error is given by the form of the error function (in our case $\varepsilon \in [0,1]$). Then, we fit the remaining data points (grey solid line, 'fit') with the error model of Eq.~\eqref{eq:error-model-decaychannel}, as described in the text, with $a_1=0.51$, $b_1=0.039$ and $c_1=0.098$.}
   \label{fig:Omega-scan-decaychannel}
\end{figure}

Here, our aim is to optimize the state transfer between a random initial state and a random target state (both taken as pure), and to investigate 100 instances of such random pairs of pure states. For this purpose, each random instance is optimized with dCRAB for different values of the bandwidth $\Delta\Omega$. Fig.~\ref{fig:Omega-scan-decaychannel} shows the scaling of the maximum control error (maximized over the random instances of the initial and target states) with the bandwidth.
Again, we consider two models for the scaling 
of the control error $\varepsilon$. Specifically, from Eq.~\eqref{eq:error-limit-wn}, one can employ the model
\begin{eqnarray}\label{eq:error-model-decaychannel}
\varepsilon_1 \gtrsim a_1\exp(-b_1\Delta\Omega)+c_1
\end{eqnarray}
up to some constants $a_1$, $b_1$ and $c_1$. To verify this model from the numerical data, 
the free parameters of the model are fit by excluding the data point in correspondence of the smallest value of $\Delta\Omega$ that can be considered as an outlier. Indeed, for a small bandwidth of the control pulse (virtually a constant pulse), 
the control error $\varepsilon$ takes the maximal possible value $1$. The same procedure is carried out also for the alternative model $\varepsilon_2= a_2 (\Delta\Omega)^{-b_2}+c_2$. For the fit of $\varepsilon_1$, shown in Fig.~\ref{fig:Omega-scan-decaychannel}, we obtain a RMS of $0.017$, while for $\varepsilon_2$ the RMS is around $0.063$. We can thus conclude that the numerical data favors an exponential scaling of the control error with the bandwidth, as opposed to a power-law scaling in agreement with Eq.\,\eqref{eq:error-limit-wn}.

As a remark, let us note here that the bandwidth $\Delta\Omega$ plays a quite crucial role. Indeed, in this case, the optimal control strategy has to move the quantum system as fast as possible into the decoherence-free subspace (or more precisely in the non-decaying state $|0\rangle\!\langle 0|$), by then giving the system another fast kick, right before the end of the evolution time, to steer it into the target state. The higher the admissible bandwidth is, the faster these kicks can be, and so the smaller the time the quantum system is exposed to decay and the smaller the control error. For this reason, in this numerical example we can observe the bandwidth-dependence of the control error as predicted by the information-theoretical bound for Gaussian white noise, Eq.~\eqref{eq:error-limit-wn}.

\section{Examples with Gaussian colored noise}\label{sec:IX}

\begin{figure}[t]
	\centering
	\includegraphics[width=0.48\textwidth]{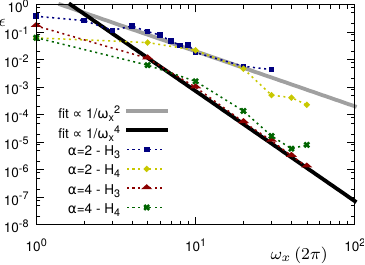}
	\caption{State protection (Sec.~\ref{sec:state-protection}): Control error scaling with respect to the Rabi frequency $\omega_x$ of the driving field for $1/f^2$- and $1/f^4$-noise and state preservation in $\frac{|0\rangle+|1\rangle}{\sqrt{2}}$. For both types of noise, we investigated separately the evolution of the quantum system according to the Hamiltonians $H_3$ and $H_4$. The Hamiltonian $H_3$ includes only a noise and a control term, and in such a scenario just a constant pulse already encodes all the information to carry out the control problem. We can observe that the control error scales with the expected $1/\omega_x^\alpha$ law provided by both the information theoretical bound and the analytic decoherence formula. In the second scenario the dynamics is governed by $H_4$ that includes also static terms in the Hamiltonian; thus, a time-modulated control pulse is needed. Yet, the scaling with $\omega_x^\alpha$ is still roughly maintained.}
	\label{fig:simulation-Kurizki-formula}
\end{figure}
To test the analytical results in section~\ref{sec:KK-formula}, we simulate both the quantum dynamics of one qubit governed by the Hamiltonian
\begin{eqnarray}
H_3(t) \equiv f(t)\sigma_x + \xi(t)\sigma_z \,,
\end{eqnarray}
where $\xi(t)$ is a colored noise field with spectrum $\varphi_{\xi}(\omega)$, and the quantum dynamics generated by the Hamiltonian
\begin{eqnarray}\label{eq:H2}
H_4(t) \equiv f(t)\sigma_x + \omega_y \sigma_y + (\omega_z + \xi(t))\sigma_z \,.
\end{eqnarray}
For both cases, in the numerical simulations, we set $T=1$, $\omega_y = \omega_z = 2\pi$, and we ensure that $\int_0^T |f(t)|^2 dt = \omega_x^2$ and $\int_0^T |\xi(t)|^2 dt = 4\pi^2$ for all noise realizations (power constraints). Moreover, for the stochastic noise field, the $1/f^\alpha$-noise with $\alpha\in\{2,4\}$ is considered. To simulate the resulting quantum dynamics, we have generated 20 realizations of each noise field and solved the SSE for each realization separately. The control objectives are then calculated as the average fidelity from all the 20 realizations.

\subsection{State protection}\label{sec:state-protection}

First, let us consider as initial state the density operator $\rho(0) = \frac{1}{2}\left(|0\rangle + |1\rangle\right)\left(\langle 0| + \langle 1|\right)$ with the aim to preserve the system in this state (i.e., $\rho_T=\rho(0)$). For the scenario $H_3$ (i.e., for the quantum dynamics described by the Hamiltonian $H_3(t)$) we keep the pulse $f(t)=\omega_x=\mathrm{const}$. Instead, for the scenario $H_4$ (i.e., for the quantum dynamics originated by the Hamiltonian $H_4(t)$) the aim is to find the optimal solution $f(t)$ that respects the power constraints, with a pulse modulation corresponding to $D = 10$.
From section~\ref{sec:KK-formula} we expect that $\omega_x$ determines the support of the filter function (i.e., $\omega_c\approx \omega_x$), such that the control error $\varepsilon$ follows the power-law behaviour corresponding to the noise coefficient $\alpha$, i.e., $\varepsilon \gtrsim 1/\omega_x^\alpha$. Fig.~\ref{fig:simulation-Kurizki-formula} shows the numerical results that confirm the expected scaling of the control error as a function of $\omega_x$. The numerical data obtained for the case of $H_s(t) = H_3(t)$ are much better described by theory (where no control resources have to be used to counter the drift term in the Hamiltonian) than the data with $H_s(t) = H_4(t)$. However, also the results for the scenario $H_4$ confirm the theoretical scaling, although here the control field has also to counter the additional drift terms in the Hamiltonian. Thus, the filter function is potentially spectrally broader and the approximations of section \ref{sec:KK-formula} become less precise.
To quantify the correspondence between numerical data and simulation, we employ the model $\log(\varepsilon)=a + b\log(\omega_x)$ that is fit to the logarithm of the numerical control error, where we treat again the smallest value of $\omega_x$ as an outlier. 
From this fit we obtain the parameter $b$, whose value is our estimate for $\alpha$. Specifically, we find 
$b=1.7\pm 0.1$ for ($\alpha=2$ and $H_3$),
$b=2.4\pm 0.3$ for ($\alpha=2$ and $H_4$),
$b=4.0\pm 0.1$ for ($\alpha=4$ and $H_3$), and
$b=3.3\pm 0.3$ for ($\alpha=4$ and $H_4$). 
Summarizing, we can thus clearly distinguish the $1/f^2$-noise from the $1/f^4$-noise, with a precision of about $0.1-0.3$ for the scenario $H_3$ and slightly smaller for $H_4$ in agreement with the bounds derived in Sec.~\ref{sec:KK-formula} from the information-theoretical aproach as well as from the filter-function approach for the example with $1/f$-noise.

\subsection{State transfer}\label{sec:state-transfer}

\begin{figure}[t]
 \centering
 \includegraphics[width=0.48\textwidth]{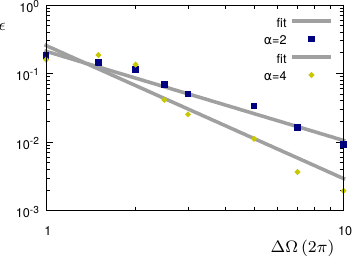}
 \caption{State transfer (Sec.~\ref{sec:state-transfer}): Control error scaling with respect to the bandwidth $\Delta\Omega$ of the control pulse for the $1/f^2$- and $1/f^4$-noise and state transfer between two random states. The time evolution is given by the Hamiltonian $H_4$. For each pair of $\alpha$ and $\Delta\Omega$, the figure shows the maximum  control error over all pairs of random initial and target states. For $\alpha\in\{2,4\}$ we fit the scaling of the control error $\varepsilon \propto 1/(\Delta \Omega)^a$, whereby we have obtained $a=1.3$ and $a=1.9$ for $\alpha=2$ and $\alpha=4$, respectively.}
 \label{fig:simulation-oneoverf4}
\end{figure}
Then, we consider a more general case, where the initial and target states are chosen randomly. We thus examine 50 different random pairs of initial and target states for each set of investigated parameters; we choose the scenario $H_4$ and we fix the power of the control pulse by setting $\omega_x=2\pi\times 5$. Our aim is to find the optimal choice of the control pulse that minimizes the error of the control problem for different values of the control bandwidth and for $1/f^2$- and $1/f^4$-noise. In Fig.~\ref{fig:simulation-oneoverf4} we show the numerical results for the maximum error over all the instances of random initial and target states for both types of noise and each value of the bandwidth $\Delta\Omega\in 2\pi[1,10]$. We expect that the control error scales as $\varepsilon \propto 1/(\Delta \Omega)^a$. Numerically, we find instead $a=1.3$ and $a=1.9$ for $\alpha=2$ and $\alpha=4$, respectively. Thus, due to the more complex dynamics, we do not recover exactly the scaling as in Fig.~\ref{fig:simulation-Kurizki-formula} or a similar analytic prediction. However, we still find a clear difference in the error scaling for the two types of noise, since the fast modulations of the system suppress the low-frequency noise more efficiently than the high-frequency noise. As a consequence, we can conclude that the control resources are used to operate the quantum system in a regime where it is less affected by the noise. In particular, if the noise power spectral density is decreasing with increasing frequency, increasing the bandwidth of the control (and thus its information content) allows 
to decrease the control error. Moreover, if the noise decreases faster, also the error scaling is more advantageous, and less resources are required to obtain the same control error compared to a slowly decreasing noise power spectral density.

\section{Conclusions}

We have studied the error scaling of a quantum control problem as a function of the noise level and control resources from an information theoretical perspective, as well as from a dynamic perspective both analytically and numerically. We have shown that the two approaches lead to the same scaling and thus are consistent. To achieve the results we have extended the information theoretical model of Ref.~\cite{Lloyd1} to colored noise and we have performed numerical examples for both white noise and colored noise to investigate the relationship between the information theoretical error bounds with the results of optimal control solutions. Furthermore, we have also analytically investigated the dynamics of a quantum system coupled to a temporally correlated environment, with the help of the Kofman-Kurizki universal formula, and then recovered the same scaling of the control error with the control modulation as predicted by the information theoretical bounds. A generalization of these results to the case of quantum gate optimization seems possible by replacing the dimension of the reachable states with the dimension of the special unitary group containing the gates. This could be investigated in a future work.

We expect that our results can pave the way towards an improved design of optimized control pulses for open quantum systems, whereby the optimization is performed also with respect to some environmental features. High fidelity control of open quantum systems, indeed, is an extremely important ingredient in the emerging field of quantum technologies, where one needs to perform precise operations under realistic (noisy) settings.

\section*{Acknowledgments}

M.M.M and T.C. acknowledge funding from the European Union’s Horizon 2020 research and innovation programme under Grant Agreement No.\,817482 (PASQuanS), as well as from the Deutsche Forschungsgemeinschaft (DFG, German Research Foundation) under Germany's Excellence Strategy -- Cluster of Excellence Matter and Light for Quantum Computing (ML4Q) EXC 2004/1 -- 390534769 and from the German Federal Ministry of Education and Research (BMBF Project No. 13N16210, SPINNING).
S.G. and F.C. acknowledge funding from the European Union’s Horizon 2020 research and innovation programme under Grant Agreement No.\,828946 (PATHOS), and from University of Florence through the project Q-CODYCES.
S.G. acknowledges The Blanceflor Foundation for financial support through the project ``The theRmodynamics behInd thE meaSuremenT postulate of quantum mEchanics (TRIESTE)''.
S.M. acknowledges support from the European Union's Horizon 2020 research and innovation programme under the Marie Skłodowska-Curie grant agreement No.\,765267 (QuSCo), and No.\,817482 (PASQuanS), by the Italian PRIN 2017 and the CARIPARO project QUASAR.

\section*{Data availability}
The datasets used for this article are available from the corresponding author on reasonable request.\\

\appendix
\section{Derivation of the master equation Eq.~(10)}
Let us start the derivation from Eq.~(8) in the main text, i.e.,
\begin{eqnarray*}
	&\dot{\rho}(t) = -i[H_s(t),\rho(t)]& \nonumber \\
	& \displaystyle{ - i\bigg[\xi(t)H_n,\left(\rho(0)-i\int_0^t [H(t'),\rho(t')] dt'\right)\bigg] }.&
\end{eqnarray*}
After some straightforward calculations, one ends-up to the following relation:
\begin{eqnarray*}
	&\dot{\rho}(t) = -i[H_s(t),\rho(t)]& \nonumber \\
	&\displaystyle{-\xi(t)\left\{i\,[H_n,\rho(0)] + \int_{0}^{t}\Big[H_n,[H_s(t'),\rho(t')]\Big]dt'\right\}}& \nonumber \\
	&-\left[H_n,\Big[H_n,\displaystyle{\int_{0}^{t}\xi(t)\xi(t')\rho(t')dt'}\Big]\right].&
\end{eqnarray*}
In this way, by averaging over the noise realizations and using the assumption $\langle \xi(t)\rangle = 0$, one gets
\begin{eqnarray*}
	&\langle\dot{\rho}(t)\rangle = -i[H_s(t),\langle\rho(t)\rangle]& \nonumber \\
	&-\left[H_n,\Big[H_n,\displaystyle{\int_{0}^{t}\langle\xi(t)\xi(t')\rho(t')\rangle dt'}\Big]\right],&
\end{eqnarray*}
where 
\begin{eqnarray*}
	&& \langle f_1(t)\rangle \equiv \int_{f_1} p_{t}(f_1)f_1 \, df_1  \\
	&&\langle f_2(t)f_3(t')\rangle \equiv \int_{f_2}\int_{f_3} p_{t,t'}(f_2,f_3)f_2 f_3 \, df_2 \, df_3 
\end{eqnarray*}
with $f_1$, $f_2$, $f_3$ here representing generic stochastic processes. Accordingly, Eq.~(10) in the main text is recovered under the further assumption to consider $\xi$ and $\rho$ as \emph{uncorrelated} processes, so that the following approximation holds:
\begin{eqnarray*}
	\int_0^t \langle\xi(t)\xi(t')\rho(t')\rangle dt' &\approx& \int_0^t \langle\xi(t)\xi(t')\rangle\langle\rho(t')\rangle dt' \nonumber \\
	&=& \int_0^t R_{\xi}(t,t')\langle\rho(t')\rangle dt'.
\end{eqnarray*}


\end{document}